\documentclass[10pt,superscriptaddress,prl,aps,longbibliography,twocolumn,floatfix]{revtex4-2}
\pdfoutput=1

\usepackage{graphicx}
\usepackage{dcolumn}
\usepackage{bm}
\usepackage{amsmath,amssymb,amsfonts,amsthm}
\usepackage{array}
\usepackage{dsfont}
\usepackage{xcolor}

\usepackage{tikz}
\usetikzlibrary{shapes,arrows,positioning,automata,backgrounds,calc,er,patterns}
\usepackage{todonotes}
\usepackage[final]{changes}
\newcolumntype{P}[1]{>{\centering\arraybackslash}p{#1}}
\usepackage[colorlinks=true,urlcolor=blue,citecolor=blue,allcolors=teal]{hyperref}
\usepackage{cleveref}
\usepackage{multirow}
\usepackage{amsmath,amssymb,bm}
\usepackage{mathtools}
\usepackage{physics}
\usepackage{enumitem}
\newcommand{\etaemb}{\eta_{\rm emb}}
\newcommand{\CE}{\mathcal C_{\rm E}}
\newcommand{\DNL}{\mathcal D^{\rm NL}}

\begin{document}

% \title{Generation of universal entanglement embezzlement resource and diverging non-local magic under generic local quantum chaotic evolution}

\title{Universal entanglement embezzlement and divergent nonlocal magic from generic local chaotic quantum evolution}

\newcommand{\Aalto}{Department of Applied Physics, Aalto University, FI-00076 Aalto, Espoo, Finland}
\newcommand{\ICMM}{Interdisciplinary Centre for Mathematical Modelling and Department of Mathematical Sciences,\\ Loughborough University, Loughborough, Leicestershire LE11 3TU, United Kingdom}
\newcommand{\LUPhys}{Department of Physics, Loughborough University, Loughborough, LE11 3TU, United Kingdom}
\newcommand{\Tampere}{Computational Physics Laboratory, Physics Unit, Faculty of Engineering and Natural Sciences, Tampere University, P.O. Box 692, FI-33014 Tampere, Finland}
\newcommand{\Helsinki}{Helsinki Institute of Physics P.O. Box 64, FI-00014, Finland}

\author{Matias Karjula}
\affiliation{\Aalto}
\author{Teemu Ojanen}
\email[\vspace{-3pt}]{teemu.ojanen@tuni.fi}
\affiliation{\Tampere}
\affiliation{\Helsinki}
\author{Kim P\"oyh\"onen}
\affiliation{\Tampere}
\author{Tapio Ala-Nissila}
\email[\vspace{-3pt}]{tapio.ala-nissila@aalto.fi}
\affiliation{\Aalto}
\affiliation{\ICMM}
\author{Moein N. Ivaki}
\email[\vspace{-3pt}]{moein.najafiivaki@aalto.fi}
\affiliation{\Aalto}

\begin{abstract}
We show that, starting from a product state, local unitary quantum evolutions generate intermediate states which exhibit a multiscale entanglement-spectrum structure required for universal entanglement embezzlement. This facilitates entanglement extraction from a catalyst many-body state while leaving it asymptotically unchanged. Remarkably, these atypical structures emerge generically at intermediate stages, well before reaching maximum entropy where thermalization has flattened out the spectral hierarchy. The resulting state is accompanied by nonlocal nonstabilizerness that diverges with the system size, consistent with a recently established equivalence between universal embezzlement and divergent nonlocal magic. Thus, without any fine tuning, a chaotic quantum evolution generates intermediate states which form a universal family of catalytic reservoirs. 
\end{abstract}
\maketitle

\emph{\textbf{Introduction.}} Entanglement is an indispensable resource of quantum information processing~\cite{RevModPhys.80.517,popescu2006entanglement}. However, it is well known that the amount of entanglement alone is not sufficient to determine resourcefulness of quantum states~\cite{gottesman1998heisenberg}. Under local operations and classical communication (LOCC), which are the free operations of the entanglement resource theory~\cite{RevModPhys.91.025001}, the possibility of transforming one bipartite pure state into another is controlled by the full set of Schmidt coefficients. This observation becomes particularly striking in the presence of catalysts~\cite{RevModPhys.96.025005,PhysRevLett.83.3566}, as auxiliary entangled states can enable transformations that would otherwise be forbidden, while being returned essentially unchanged. An extreme manifestation is \emph{entanglement embezzlement}~\cite{PhysRevA.67.060302,van2024embezzlement, PhysRevA.90.042331,van2025multipartite,pfyl-hwf2, PhysRevA.62.012304}, in which an appropriate family of catalyst states can supply, to arbitrarily high accuracy, essentially any desired bipartite entangled state while undergoing an asymptotically vanishing disturbance. Such states therefore behave as universal reservoirs of entanglement in the thermodynamic limit, i.e., they form an entanglement battery.

Universal embezzlement (UE) requires a particular \textit{scale-free} organization of the subsystem Schmidt spectrum, and appreciable spectral weight must remain distributed over progressively separated scales as the Hilbert-space dimension grows~\cite{zanoni2024complete}. A recent result~\cite{sierant2026exact} establishes an unexpected connection to \emph{nonlocal magic}, the component of nonstabilizerness that cannot be removed by local basis transformations~\cite{gottesman1998heisenberg,PhysRevA.70.052328,PhysRevA.71.022316,torre2026non,z3vr-w5c5, huang2026intrinsic,viscardi2026non, liu2026nonlocal, liu2026entanglement}. For pure bipartite states, this quantity can be determined directly from the entanglement spectrum, and a family of states is a universal embezzler precisely when its nonlocal magic diverges in the thermodynamic limit. These therefore identify a form of quantum spectral resourcefulness tied to the presence of \textit{both entanglement and magic}, manifested in the multiscale structure of the subsystem spectrum. Despite the profound implications of this result, known realizations of UE have so far arisen in special settings. Besides the explicitly engineered harmonic catalysts~\cite{PhysRevA.67.060302}, UE appears, albeit inefficiently, in the ground-state sectors of one-dimensional critical free-fermion systems and their dual spin chains~\cite{van2025critical}. Sufficiently many copies of a fixed nonmaximally entangled bipartite state likewise form a universal embezzling family~\cite{sierant2026exact}. A broader manifestation occurs in relativistic quantum field theory, where the vacuum itself can act as a universal embezzler through a specific structure of local observable algebras~\cite{PhysRevLett.133.261602}.

To this end, the question we address here is: \emph{Can multiscale spectra arise from simple quantum states under generic local evolution?} We focus on the case of quantum chaotic systems, which can rapidly generate increasingly complex correlations~\cite{karjula2026ebbs,cotler2017chaos,haug2025probing,leone2021quantum,fisher2023random}. Although they can possess maximal coarse-grained resources, their entanglement spectra are governed by a narrow notion of statistical typicality~\cite{hayden2006aspects,mori2018thermalization,kaufman2016quantum}. It is far from obvious whether quantum chaos should generate a \textit{universal} family of embezzling states. In fact, two tendencies appear to compete. Dynamics must scramble enough to populate many spectral scales, but not so strongly that typicality erases their broad hierarchy~\cite{chang2019evolution, serbyn2016power,karjula2026ebbs,tnfv-lzfx,zhang2026revealing}. This tension need not appear in integrable or weakly thermalizing systems~\cite{collura2026nonlocal,iannotti2026non,liu2026nonlocal}. 

We show that in closed chaotic quantum systems this competition produces a distinct intermediate regime of unitary quantum  evolution. Starting from simple product states and evolving unitarily under local Haar-random dynamics, entanglement spectrum progressively spreads across logarithmic rank scales and the system develops increasing embezzling power, with the characteristic time of embezzling scaling linearly with the system size \(N\). The corresponding transient peaks of nonlocal magic and entanglement capacity diverge as \(\log_2 N\) and \(N\), respectively, while the embezzling parameter simultaneously vanishes as a power law. Continued evolution, however, eventually drives the reduced state toward regimes where the multiscale structure is lost. We find analogous behaviors in a family of parametrically tunable random circuits, where varying an interaction strength continuously interpolates between weakly entangled, strongly scrambled, and a special maximally entangled Clifford-only point. The results establish nonequilibrium local many-body evolution as a physical route from simple states to universal entanglement embezzlers (UEEs). 

\emph{\textbf{Entanglement embezzlement and nonlocal magic.}} Entanglement cannot be created by LOCC, but it can be \emph{borrowed} from a suitable catalyst while leaving the latter asymptotically unchanged. The canonical construction of van Dam and Hayden~\cite{PhysRevA.67.060302} is the family
\begin{equation}
|\Gamma_d\rangle_{AB}=
\frac{1}{\sqrt{H_d}}
\sum_{x=0}^{d-1}\frac{1}{\sqrt{x+1}}
|x\rangle_A|x\rangle_B ,
\quad
H_d=\sum_{j=1}^{d}\frac{1}{j},\nonumber
\label{eq:vdh}
\end{equation}
where \(d\) is the local catalyst dimension (equivalently, the Schmidt rank) and \({|x\rangle_{A,B}}\) are local Schmidt bases. Its Schmidt spectrum is \(\lambda_x=[(x+1)H_d]^{-1}\). For any fixed finite-dimensional bipartite target state \(|\sigma\rangle_{A'B'}\), sufficiently large members of this family permit \[\ket{\Gamma_d}_{AB}\otimes\ket{00}_{A'B'}
\xrightarrow{\hat V_{AA'}\otimes \hat V_{BB'},}
\ket{\Gamma_d}_{AB}\otimes|\sigma\rangle_{A'B'},\] where \(\hat V_{AA'}\) and \(\hat V_{BB'}\) are local unitaries acting on the two parties' respective catalyst and ancillary registers (or, equivalently, local isometries upon adjoining the ancillas), and the distance from the ideal product \(|\Gamma_d\rangle_{AB} \otimes |\sigma\rangle_{A'B'} \) vanishes as \(d\to\infty\). Remarkably, for this family local unitaries alone suffice to extract the target entanglement, through a rearrangement of the Schmidt amplitudes that leaves the catalyst asymptotically arbitrarily close in trace distance to its initial state~\cite{RevModPhys.96.025005,PhysRevLett.127.150503}. A sequence with this property for \emph{every} finite target is called a \textit{universal embezzling family}. The origin of this counterintuitive property lies in the scale-free form of the entanglement spectrum, and the fact that trace-norm closeness does not guarantee entropy closeness when the catalyst dimension becomes unbounded~\cite{zanoni2024complete}. For example, extracting a Bell pair approximately replaces the ordered Schmidt probabilities according to $\lambda_x\rightarrow\lambda_{\lfloor x/2\rfloor}/2$, corresponding to a translation along the logarithmic rank axis. For the canonical construction, this is enabled by the approximate invariance of the catalyst spectrum under such dilations.

To make this explicit, let us partition the ordered spectrum into the so-called dyadic shells~\cite{liu2026entanglement}
\begin{equation}
D_0=\{0\},\
D_j=\{2^{j-1},\ldots,2^j-1\},\
w_j=\sum_{x\in D_j}\lambda_x ,
\label{eq:octave_weights}
\end{equation}
with the last shell truncated at the Schmidt rank. \(w_j\) is essentially the probability density of Schmidt weight per unit logarithmic rank. For the van Dam--Hayden family, $w_0=H_d^{-1}$ and, for $j\geq1$, \(w_j\simeq\ln 2/H_d\), giving equal Schmidt weight in every octave of rank. This admits a familiar mathematical interpretation as the canonical spectrum is a finite-size Zipf law~\cite{PhysRevLett.113.068102,mora2011biological}. Its normalization $H_d$ is the truncated Dirichlet series associated with the pole of the Riemann zeta function \(\zeta(s)\!=\!\sum_{j=1}^{\infty}j^{-s}\) at $s\!=\!1$, \(\zeta(1)\!=\!\infty\), yielding $H_d\!=\!\ln d+\gamma+o(1)$, with \(\gamma\) a known constant. Unlike steeper power laws, whose normalization approaches $\zeta(s)<\infty$ for $s>1$, the marginal exponent $s=1$ distributes a finite fraction of the total Schmidt weight over an ever-growing number of logarithmic rank scales~\footnote{Equivalently, any fixed multiplicative interval $[x,cx]$ carries asymptotically rank-independent weight $\sim\ln(c)/H_d$, explaining the nearly uniform dyadic weights.}. Within regular power-law families~\cite{liu2026entanglement}, UE occurs only at $\lambda_x\!\sim\! x^{-1}$, while more general universal families, such as \(\lambda_x\!\propto\! f(x)/x\), with \(f(x)\) a slow varying function, can accommodate corrections to this scaling~\cite{PhysRevA.90.042331}. 

More generally, allowing LOCC, a practical criterion is given by the largest weight contained in any factor-of-two window,
\begin{equation}
\etaemb=\max_{\ell\geq0}
\sum_{x=\ell}^{2\ell}\lambda_x ,
\label{eq:eta_emb}
\end{equation}
where the spectrum is zero padded beyond its Schmidt rank, and \(\max_jw_j\leq\etaemb\leq2\max_jw_j\). A family $\{|\Gamma_d\rangle\}$ is a universal LOCC embezzler iff $\etaemb(\Gamma_d)\!\rightarrow\!0$ as $d\!\rightarrow\!\infty$, as established in Ref.~\cite{zanoni2024complete}. For the canonical family, $\etaemb\!=\!H_d^{-1}\!\sim\!1/\ln d$ exactly, and \(\etaemb\sqrt{\CE}\!\to\!(\sqrt{12}\ln2)^{-1}\), where \(\CE\!=\!\operatorname{Var}_{\lambda}(-\log_2\lambda)\) is the capacity of entanglement~\cite{PhysRevD.99.066012}.
\begin{figure*}[t]
\centering
\includegraphics[width=0.99\linewidth]{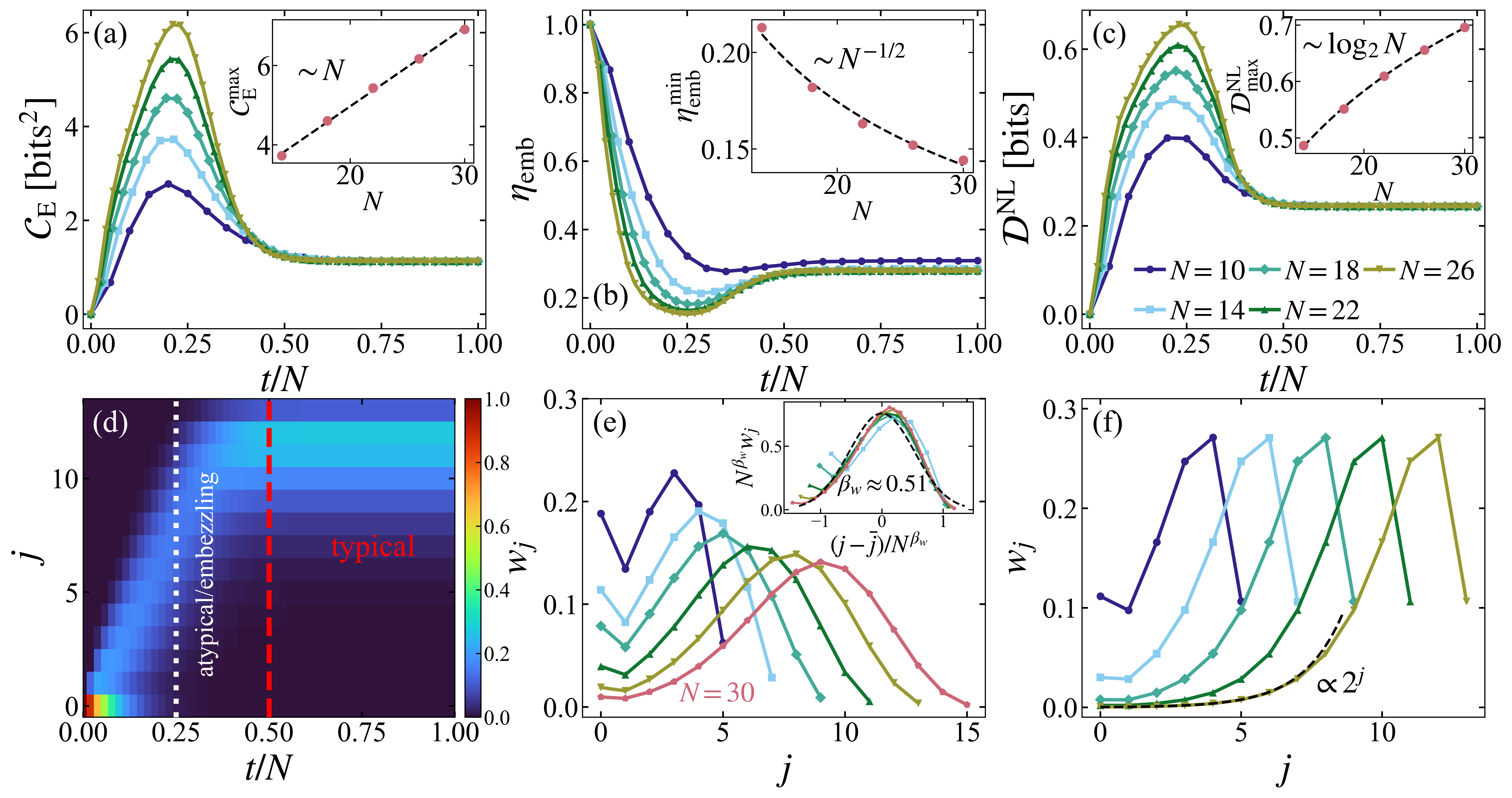}
\caption{Dynamical creation and destruction of embezzling capabilities in the Haar-random model. At their corresponding extrema, the ensemble-averaged quantities scale as \textbf{(a)} \(\mathbb{E}_U[\CE]\propto N\),
\textbf{(b)} \(\mathbb{E}_U[\etaemb]\propto N^{-\beta_{\eta}}\) with \(\beta_{\eta}\approx1/2\), and
\textbf{(c)} \(\mathbb{E}_U[\DNL]\propto \log_2 N\). Further, \(\mathcal D^{\rm NL, Haar}\approx0.244,\ \mathcal{C}^{\rm Haar}_{\rm E}\approx1.124\), and \(\eta^{\rm Haar}_{\rm emb}\approx0.278\). The dotted and dashed lines in \textbf{(d)} denote the atypical embezzling point and the crossover to the typical regime, respectively, for \(N\!=\!26\) qubits. \textbf{(e)} and \textbf{(f)} display the average dyadic shell distributions \(\mathbb{E}_U[w_j]\) for the optimal embezzling and Haar-typical points, respectively. Results are averages over \(50\)--\(400\) independent realizations.}
\label{fig:haar}
\end{figure*}
As noted, a recent result~\cite{sierant2026exact} reveals that the same multiscale spectral structure controls nonlocal magic. For a pure bipartite state with ordered Schmidt probabilities $\{\lambda_x\}$, the nonlocal min-relative entropy of nonstabilizerness is
\begin{equation}
\DNL=-\log_2
\max_{0\leq k\leq\nu}
\left[\frac{1}{2^k}
\left(\sum_{x=0}^{2^k-1}\sqrt{\lambda_x}
\right)^2\right],
\label{eq:dminnl}
\end{equation}
where $\nu=\min(N_A,N_B)$. This admits a simple interpretation. Stabilizer states possess, up to local unitaries, flat entanglement spectra of dyadic rank $2^k$, corresponding to $k$ Bell pairs, and $\DNL$ measures the distance of the actual spectrum from the closest such sector. Crucially, defining \(\mu_{\rm emb}\!=\!-\log_2\etaemb\), it is further shown that \( \mu_{\rm emb}-\log_2(3+2\sqrt2)
\!\leq\! \DNL\!\leq\!\mu_{\rm emb}.\) Hence, a family of states becomes a UEE precisely when nonlocal magic diverges. These are therefore characterizations of the same scale-delocalized organization of the entanglement spectrum. Crucially, for balanced cuts, nonlocal stabilizer R\'enyi entropy \(\mathcal{M}^{\rm NL}_{\rm SRE}\)~\cite{PhysRevLett.128.050402, z3vr-w5c5, liu2026nonlocal} is known to obey the bounds $\DNL/5\leq \mathcal{M}_{\rm SRE}^{\rm NL}\leq4\DNL$~\cite{sierant2026exact}, so it diverges for precisely the same state families. The Schmidt-gauged nonlocal magic~\cite{franchini2026schmidt} is also expected to inherit the same behavior. Moreover, one has $0\leq \DNL\leq\log_2(\nu+1)$, which is the fastest asymptotic growth allowed by the dyadic geometry. We will exploit these and show a random quantum evolution can dynamically create, and subsequently destroy, this structure.

\emph{\textbf{Haar-random circuits.}} We first consider a paradigmatic model of local quantum chaotic evolution; that is, a one-dimensional brickwork circuit composed of independently sampled two-qubit Haar-random unitaries~\cite{fisher2023random}. For a ring of $N$ qubits, one time step consists of even-odd alternating layers \(
\hat U(t)=\hat U_{\rm o}(t)\hat U_{\rm e}(t),\) with \(\hat U_{\rm e/o}(t)=\prod_{(i,j)\in{\rm e/o}} \hat U_{ij}(t)\) where each $\hat U_{ij}(t)$ is drawn independently from the Haar measure on $U(4)$. We initialize the system in a product state, $\ket{0}^{\otimes N}$, and consider an equal complementary and contiguous bipartition $A|A'$. Local scrambling produces ballistic entanglement growth, so that the half-system entropy reaches its volume-law, random-state value at a depth $t_{\rm ent}\simeq N/2$~\cite{nahum2017quantum}. 
\begin{figure*}[t]
\centering
\includegraphics[width=0.99\linewidth]{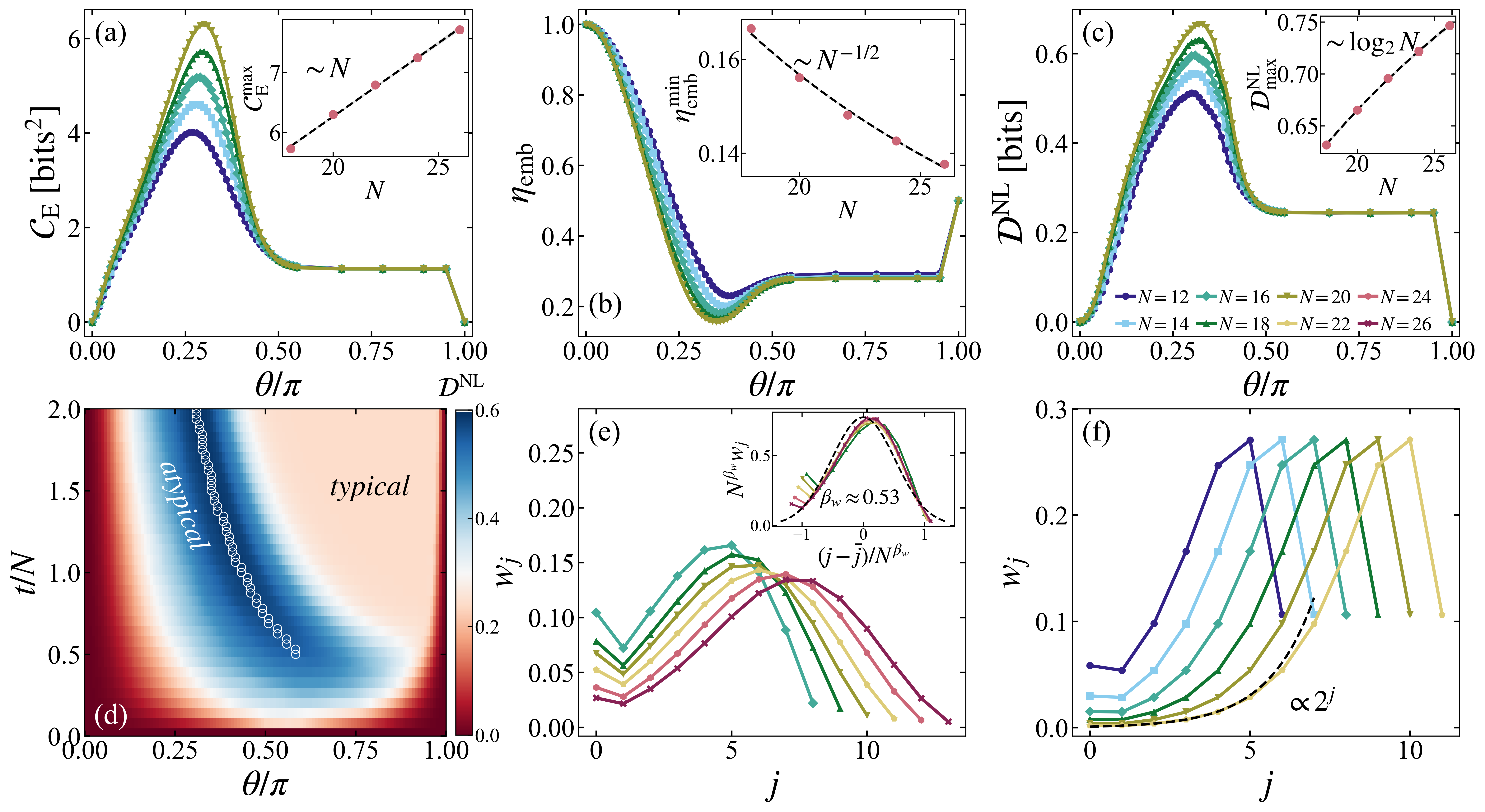}
\caption{Parametric creation and destruction of the scale-free embezzling structure in a tunable random circuit model. Quantities in \textbf{(a)}, \textbf{(b)}, \textbf{(c)} and their overall scaling behavior are consistent with those of the Haar-random model. Plotted for a fixed depth \(t\!=\!2N\). In \textbf{(d)} we further plot the resource landscape displaying a parametric-dynamical ridge, along which nonlocal magic and embezzling power are maximized. The hollow circles denote
\(\theta_{\rm emb}(t/N)\!=\!\arg\max_\theta \mathbb{E}[\mathcal{D}^{\rm NL}(\theta,t/N)]\) for \(t/N\geq0.5\), which roughly separate the weakly structured and the strongly scrambled regimes. \textbf{(e)} and \textbf{(f)} show the shell profile for the embezzling point \(\theta/\pi\approx0.35\) and the typical point \(\theta/\pi\approx0.8\), respectively. The fitted exponents \(\beta_{\eta}\approx\beta_{w}\approx1/2\) are consistent within uncertainty and agree with the Haar-random circuit. Results are averages over \(50\)--\(200\) independent realizations.}
\label{fig:cp}
\end{figure*}
 
We resolve the dynamics through the dyadic weights $w_j$ of Eq.~\eqref{eq:octave_weights}, together with the embezzlement parameter $\etaemb$ of Eq.~\eqref{eq:eta_emb}, the entanglement capacity \(\CE\), and the associated nonlocal magic measure \(\DNL\) of Eq.~\eqref{eq:dminnl}. As shown in {\color{brown}Figure}~\ref{fig:haar}, the quantities exhibit a common nonmonotonic behavior, with an intermediate extremum~\cite{karjula2026ebbs,aditya2025growth}. The characteristic resource extrema occur at times proportional to the system size. For the largest accessible systems, the minimum of the ensemble averaged \(\etaemb\) occurs for \(t_{\rm emb}\simeq N/4\), where \(t_{\rm emb}\!=\!\arg\min_t\mathbb{E}_{U}[\etaemb(N,t)]\), and \(\mathbb{E}_{U}\) denotes averages over circuit realizations~\footnote{Given \(\mathbb{E}_U[\eta_{\rm emb}]\sim N^{-\beta_{\eta}}\) and \(\eta_{\rm emb}\geq0\), from the Markov inequality it follows that \(\rm{Pr}(\eta_{\rm emb}>\epsilon)\leq \mathbb{E}_U[\eta_{\rm emb}]/\epsilon={\mathcal{O}}(N^{-\beta_{\eta}})\)}. This defines a dynamically generated family of increasingly powerful entanglement catalysts, \(\{\ket{\Psi_{N}(t_{\rm emb})}\}_N\). The maxima of \(\mathbb{E}_{U}[\DNL]\) and \(\mathbb{E}_{U}[\CE]\) occur at nearby, generally distinct times within the same intermediate regime~\cite{zhang2026entanglement}. At around the atypical embezzling point, for the average quantities one finds \(\mathbb{E}_{U}[\etaemb]\propto N^{-\beta_{\eta}}\), \(\mathbb{E}_{U}[\DNL]\propto\log_2 N\), and \(\mathbb{E}_{U}[\CE]\propto N\). The scalings behaviors establish the possibility of asymptotic UE, with the resulting Schmidt spectra distinct from the canonical family. The \(w_j\) heatmap, {\color{brown}Figure.}~\ref{fig:haar}(d), also reveals a transient redistribution of octave weights, forming the intermediate embezzling regime. At larger depths, this multiscale structure collapses toward the Marchenko--Pastur (MP) profile, where the weight is concentrated within a finite range of octaves around the bulk Schmidt scale.

The transport-like features of \(w_j\) further sharply distinguishes the intermediate and late-time behaviors. At around the embezzling point, \(w_j\) forms a packet in logarithmic rank \(j\) that shifts, broadens, and decreases in height with increasing system size. We find that the resulting structure is well described by the finite-size traveling-scaling form
\[\mathbb{E}_U[w_j(N)]\simeq N^{-\beta_w}f\!\left(\frac{j-\bar j}{N^{\beta_w}}\right),\quad\bar j=aN+b,\]
with \(\bar j\simeq 0.32N+\mathcal{O}(1)\) and \(\beta_w\approx1/2\). This gives an approximate description of a probability envelope that is transported ballistically in its center across Schmidt scales, while simultaneously broadening diffusively in logarithmic-rank space. Under this rescaling, different system sizes collapse onto a common, near-Gaussian profile (cf. {\color{brown}Figure.} 1(e)). Thus, the octave weight spreads over a parametrically growing number \(\Delta j\sim N^{\beta_w}\) of logarithmic rank scales. Since \(\sum_j\mathbb{E}_{U}[w_j]=1\), this gives \(\max_j\mathbb{E}_{U}[w_j]\sim N^{-\beta_w}\). The same finite-size scaling is observed for \(\mathbb{E}_{U}[\etaemb]\), with
\(\beta_w\approx\beta_\eta\) within the accessible system sizes. Thus, the broadening of the spectral packet is accompanied by a vanishing characteristic shell weight and embezzlement error. The finite envelope of the packet, however, prevents a globally harmonic spectrum, and with the packet variance \(\propto\! N\), one roughly obtains \(\CE\propto N\), rather than the \(\propto N^2\) scaling of the canonical family spanning \(\mathcal{O}(N)\) logarithmic rank shells. By the bounds above, UE requires only a vanishing \(\max_j w_j\), and not an exactly Gaussian packet or a global Zipf law~\footnote{Roughly, the embezzling regime develops a broadening packet in logarithmic Schmidt rank whose central region becomes locally Zipf-like, \(\lambda_x\!\sim\! x^{-1}\), as quantified by \(\alpha_j\!=\!1-\log_2(w_{j+1}/w_j)\!\to\!1\). This conclusion is unchanged by subleading corrections to the near-Gaussian profile, such as skewness, provided they do not alter the scaling of its width. A more detailed characterization is not essential to the embezzlement argument.}.

In the Haar-typical regime, the upper Schmidt spectrum becomes approximately flat over rank windows \(r\ll d_A\), as follows from the behavior of the MP distribution~\cite{sommers2004statistical,vivo2016random}. Since the corresponding Schmidt weights are all of order \(1/d_A\), the weight contained in a dyadic shell initially grows simply with the number of Schmidt values it contains, \(w_j\simeq x_+2^{j-1}/d_A\), where \(x_+\) is the upper edge of the rescaled MP spectrum (\(x_+=4\) for an equal bipartition). This growth eventually turns over as the shell approaches the bulk of the spectrum, where the Schmidt weights decrease appreciably with rank. More generally, the entire Haar-typical shell profile follows an analytical function~\footnote{Writing \(\lambda_r\simeq d_A^{-1}x(r/d_A)\), with \(x(u)\) the upper-tail quantile of the MP distribution, one obtains \(w_j\simeq\int_{2^{j-1}/d_A}^{2^{j}/d_A}x(u)\,du\). For an equal bipartition, \(x(u)\!=\!4\cos^2\theta(u)\), where \(u\!=\![2\theta-\sin(2\theta)]/\pi\), providing a parameter-free prediction for the full \(w_j\) profile, which fits almost perfectly to our data.}. 

\emph{\textbf{Tunable random circuits.}} The emergence of the multiscale spectral structure is not specific to the system studied above. This is now illustrated on a model with a fixed preparation depth \(t\!\sim\!\mathcal{O}(N)\), which serves as a minimally tunable brickwork circuit. A convenient choice is~\cite{karjula2026ebbs, varikuti2026impact}
\begin{equation}
\hat U_{\rm F}(\theta,t)=\hat U_{\rm o}(\theta,t)C_2(t)
\hat{U}_{\rm e}(\theta,t)\hat C_1(t),\nonumber
% \label{eq:tunable_floquet}
\end{equation}
where \(\hat C_{1,2}(t)\!=\!\bigotimes_i \hat c_i(t)\) are independent single-qubit random Clifford operations, \(\hat U_{\rm e/o}(\theta,t)\!=\!\prod_{(i,j)\in{\rm e/o}}\hat P_{ij}(\theta)\), and
\(\hat P(\theta)\!=\!{\rm diag}(1,1,1,e^{i\theta})\).
The parameter \(\theta\) controls the entangling and nonstabilizing strengths, leaving the circuit geometry and gate count unchanged.

\begin{figure}
    \centering
    \includegraphics[width=0.99\linewidth]{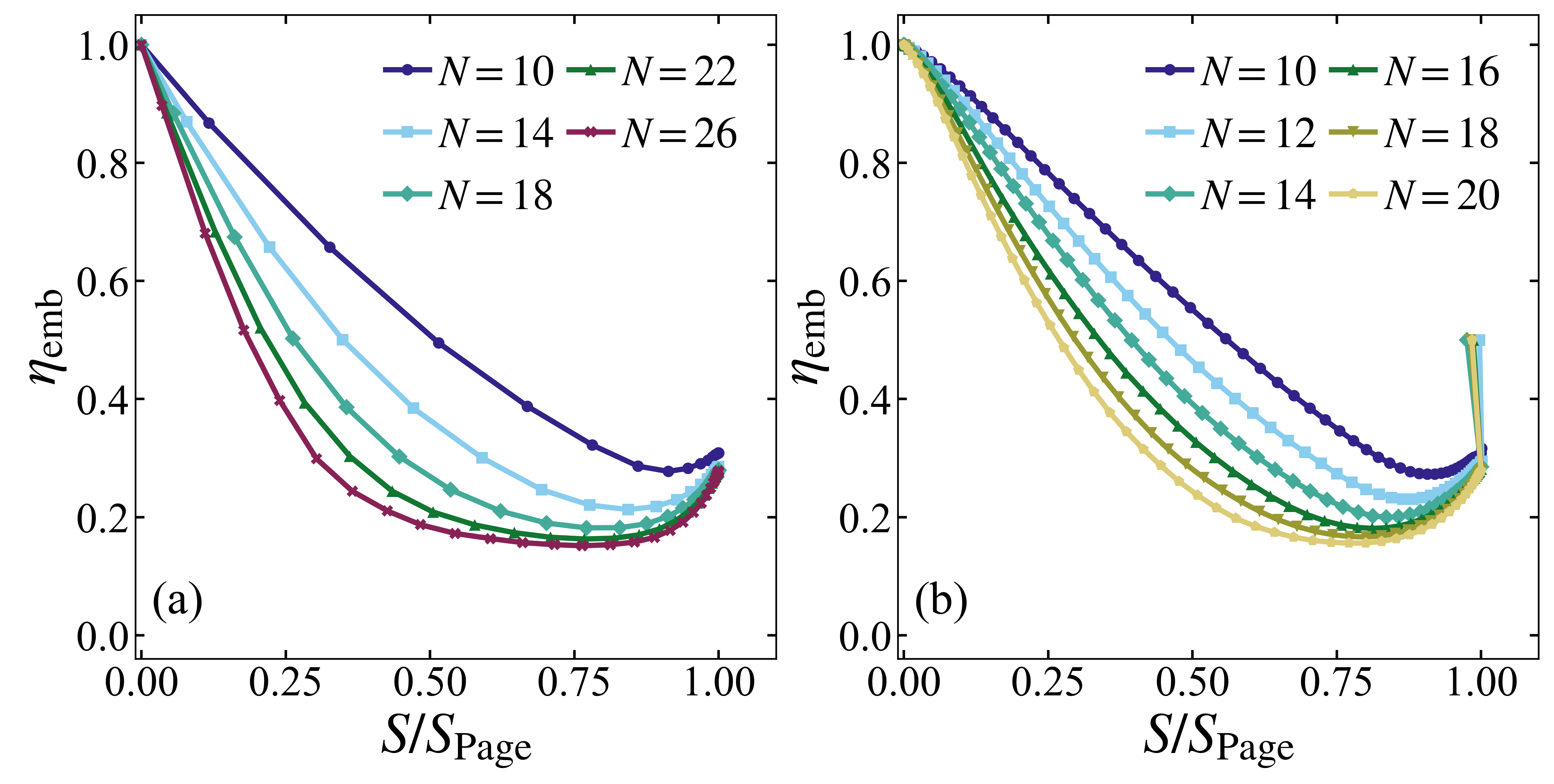}
    \caption{A flow-like diagram in the \(\etaemb\)–\(S/S_{\rm Page}\) plane, showing convergence toward a common submaximal-entanglement minimum of the embezzlement parameter with increasing system size. Shown for \textbf{(a)} the Haar-random circuit dynamics, and \textbf{(b)} the parametrically tunable random circuit.}
    \label{fig:flow}
\end{figure}

The resulting behavior, shown in {\color{brown}Figure}~\ref{fig:cp}, closely mirrors the previous case. At a fixed linear depth \(t/N=2\), increasing \(\theta\) drives the system from weakly entangled states through an intermediate regime with a broad, multiscale Schmidt spectrum and enhanced embezzling power, before strong scrambling again suppresses embezzlement. This correspondence becomes particularly transparent in the resource landscape \(\DNL(\theta/\pi,t/N)\), where the optimal embezzling states form a pronounced ridge. The ridge shifts with preparation depth, so that the phenomenon is not tied to a particular interaction strength but follows a dynamical crossover between entanglement generation and spectral randomization. At the special (maximally entangled) stabilizer point, \(\theta=\pi\), one has $D^{\rm NL}\!=\!\CE\!=\!0$, while \(\etaemb\!\neq\!0\). Indeed, for a stabilizer state with flat Schmidt spectrum \(\etaemb\!=\!1/2\). Plotted in {\color{brown}Figure}~\ref{fig:flow}, the embezzling parameter follows a nonmonotonic flow against the Page-normalized entanglement~\cite{PhysRevLett.71.1291}, reaching its asymptotic minimum at an intermediate entangling "fixed-point".

\emph{\textbf{Conclusions.}} Our results establish UEE beyond the known specially constructed families, showing that such scale-free structures are naturally generated under local chaotic manybody evolution without fine tuning. Universal-catalysis-like structure emerges in intermediate states starting from a product state on route to the maximum entropy and typicality. This highlights a distinction between the \emph{amount} of entanglement produced by manybody dynamics and its \textit{structure}. From the perspective of quantum resource theories, local chaotic evolutions generically pass through an intermediate regime that supports catalytic transformations with asymptotically vanishing error. More broadly, our work motivates asking whether manybody dynamics can similarly generate effective catalysts for resources beyond entanglement.

\emph{\textbf{Acknowledgments.}} This work was supported by the European Union and the European Innovation Council through the Horizon Europe Project No. QRC-4-ESP (Grant Agreement No. 101129663), and EU Horizon Europe Quest project (Project No. 101156088). T.O. acknowledges the support by the Finnish Research Council project 362573 and the Finnish quantum flagship program. K.P. acknowledges the support by the Finnish Research Council project 363879. This work is part of the Finnish Center of Excellence in Quantum Materials (QMAT). GPT-5.6 was used as an auxiliary tool for literature discovery, cross-checking scientific reasoning and intuitions, and assistance with code development and automation. All scientific conclusions and interpretations are those of the authors. No AI-generated prose is incorporated into the manuscript.

\bibliography{refs}

@article{van2025critical,
  title={Critical fermions are universal embezzlers},
  author={van Luijk, Lauritz and Stottmeister, Alexander and Wilming, Henrik},
  journal={Nature Physics},
  volume={21},
  number={7},
  pages={1141--1146},
  year={2025},
  publisher={Nature Publishing Group UK London}
}

@article{sierant2026exact,
  title={Exact quantification of nonlocal magic},
  author={Sierant, Piotr},
  journal={arXiv preprint arXiv:2608.28563},
  year={2026}
}

@article{PhysRevA.67.060302,
  title = {Universal entanglement transformations without communication},
  author = {van Dam, Wim and Hayden, Patrick},
  journal = {Phys. Rev. A},
  volume = {67},
  issue = {6},
  pages = {060302(R)},
  numpages = {3},
  year = {2003},
  month = {Jun},
  publisher = {American Physical Society},
  doi = {10.1103/PhysRevA.67.060302},
  url = {https://link.aps.org/doi/10.1103/PhysRevA.67.060302}}

@article{serbyn2016power,
  title={Power-law entanglement spectrum in many-body localized phases},
  author={Serbyn, Maksym and Michailidis, Alexios A and Abanin, Dmitry A and Papi{\'c}, Zlatko},
  journal={Physical review letters},
  volume={117},
  number={16},
  pages={160601},
  year={2016},
  publisher={APS}
}

@article{varikuti2026impact,
  title={Impact of Clifford operations on non-stabilizing power and quantum chaos},
  author={Varikuti, Naga Dileep and Bandyopadhyay, Soumik and Hauke, Philipp},
  journal={Quantum},
  volume={10},
  pages={2017},
  year={2026},
  publisher={Verein zur F{\"o}rderung des Open Access Publizierens in den Quantenwissenschaften}
}

@article{PhysRevLett.128.050402,
  title = {Stabilizer R\'enyi Entropy},
  author = {Leone, Lorenzo and Oliviero, Salvatore F. E. and Hamma, Alioscia},
  journal = {Phys. Rev. Lett.},
  volume = {128},
  issue = {5},
  pages = {050402},
  numpages = {5},
  year = {2022},
  month = {Feb},
  publisher = {American Physical Society},
  doi = {10.1103/PhysRevLett.128.050402},
  url = {https://link.aps.org/doi/10.1103/PhysRevLett.128.050402}
}

@article{PhysRevD.99.066012,
  title = {Aspects of capacity of entanglement},
  author = {de Boer, Jan and J\"arvel\"a, Jarkko and Keski-Vakkuri, Esko},
  journal = {Phys. Rev. D},
  volume = {99},
  issue = {6},
  pages = {066012},
  numpages = {34},
  year = {2019},
  month = {Mar},
  publisher = {American Physical Society},
  doi = {10.1103/PhysRevD.99.066012},
  url = {https://link.aps.org/doi/10.1103/PhysRevD.99.066012}
}

@article{z3vr-w5c5,
  title = {Gravitational Backreaction is Magical},
  author = {Cao, ChunJun and Cheng, Gong and Hamma, Alioscia and Leone, Lorenzo and Munizzi, William and Oliviero, Savatore F.E.},
  journal = {PRX Quantum},
  volume = {6},
  issue = {4},
  pages = {040375},
  numpages = {39},
  year = {2025},
  month = {Dec},
  publisher = {American Physical Society},
  doi = {10.1103/z3vr-w5c5},
  url = {https://link.aps.org/doi/10.1103/z3vr-w5c5}
}

@article{aditya2025growth,
  title={Growth and spreading of quantum resources under random circuit dynamics},
  author={Aditya, Sreemayee and Turkeshi, Xhek and Sierant, Piotr},
  journal={arXiv preprint arXiv:2512.14827},
  year={2025}
}

@article{cotler2017chaos,
  title={Chaos, complexity, and random matrices},
  author={Cotler, Jordan and Hunter-Jones, Nicholas and Liu, Junyu and Yoshida, Beni},
  journal={Journal of High Energy Physics},
  volume={2017},
  number={11},
  pages={1--60},
  year={2017},
  publisher={Springer}
}

@article{RevModPhys.91.025001,
  title = {Quantum resource theories},
  author = {Chitambar, Eric and Gour, Gilad},
  journal = {Rev. Mod. Phys.},
  volume = {91},
  issue = {2},
  pages = {025001},
  numpages = {48},
  year = {2019},
  month = {Apr},
  publisher = {American Physical Society},
  doi = {10.1103/RevModPhys.91.025001},
  url = {https://link.aps.org/doi/10.1103/RevModPhys.91.025001}
}

@article{gottesman1998heisenberg,
  title={The Heisenberg representation of quantum computers},
  author={Gottesman, Daniel},
  journal={arXiv preprint quant-ph/9807006},
  year={1998}
}

@article{PhysRevA.70.052328,
  title = {Improved simulation of stabilizer circuits},
  author = {Aaronson, Scott and Gottesman, Daniel},
  journal = {Phys. Rev. A},
  volume = {70},
  issue = {5},
  pages = {052328},
  numpages = {14},
  year = {2004},
  month = {Nov},
  publisher = {American Physical Society},
  doi = {10.1103/PhysRevA.70.052328},
  url = {https://link.aps.org/doi/10.1103/PhysRevA.70.052328}
}

@article{leone2021quantum,
  title={Quantum chaos is quantum},
  author={Leone, Lorenzo and Oliviero, Salvatore FE and Zhou, You and Hamma, Alioscia},
  journal={Quantum},
  volume={5},
  pages={453},
  year={2021},
  publisher={Verein zur F{\"o}rderung des Open Access Publizierens in den Quantenwissenschaften}
}

@article{haug2025probing,
  title={Probing quantum complexity via universal saturation of stabilizer entropies},
  author={Haug, Tobias and Aolita, Leandro and Kim, MS},
  journal={Quantum},
  volume={9},
  pages={1801},
  year={2025},
  publisher={Verein zur F{\"o}rderung des Open Access Publizierens in den Quantenwissenschaften}
}

@article{PhysRevLett.71.1291,
  title = {Average entropy of a subsystem},
  author = {Page, Don N.},
  journal = {Phys. Rev. Lett.},
  volume = {71},
  issue = {9},
  pages = {1291--1294},
  numpages = {0},
  year = {1993},
  month = {Aug},
  publisher = {American Physical Society},
  doi = {10.1103/PhysRevLett.71.1291},
  url = {https://link.aps.org/doi/10.1103/PhysRevLett.71.1291}
}

@article{RevModPhys.80.517,
  title = {Entanglement in many-body systems},
  author = {Amico, Luigi and Fazio, Rosario and Osterloh, Andreas and Vedral, Vlatko},
  journal = {Rev. Mod. Phys.},
  volume = {80},
  issue = {2},
  pages = {517--576},
  numpages = {0},
  year = {2008},
  month = {May},
  publisher = {American Physical Society},
  doi = {10.1103/RevModPhys.80.517},
  url = {https://link.aps.org/doi/10.1103/RevModPhys.80.517}
}

@article{mori2018thermalization,
  title={Thermalization and prethermalization in isolated quantum systems: a theoretical overview},
  author={Mori, Takashi and Ikeda, Tatsuhiko N and Kaminishi, Eriko and Ueda, Masahito},
  journal={Journal of Physics B: Atomic, Molecular and Optical Physics},
  volume={51},
  number={11},
  pages={112001},
  year={2018},
  publisher={IOP Publishing}
}

@article{tnfv-lzfx,
  title = {Optimal quantum reservoir learning in proximity to universality},
  author = {Ivaki, Moein N. and Karjula, Matias and Ala-Nissila, Tapio},
  journal = {Phys. Rev. A},
  volume = {113},
  issue = {6},
  pages = {L060401},
  numpages = {8},
  year = {2026},
  month = {Jun},
  publisher = {American Physical Society},
  doi = {10.1103/tnfv-lzfx},
  url = {https://link.aps.org/doi/10.1103/tnfv-lzfx}
}

@article{mora2011biological,
  title={Are biological systems poised at criticality?},
  author={Mora, Thierry and Bialek, William},
  journal={Journal of Statistical Physics},
  volume={144},
  number={2},
  pages={268--302},
  year={2011},
  publisher={Springer}
}

@article{PhysRevLett.113.068102,
  title = {Zipf's Law and Criticality in Multivariate Data without Fine-Tuning},
  author = {Schwab, David J. and Nemenman, Ilya and Mehta, Pankaj},
  journal = {Phys. Rev. Lett.},
  volume = {113},
  issue = {6},
  pages = {068102},
  numpages = {5},
  year = {2014},
  month = {Aug},
  publisher = {American Physical Society},
  doi = {10.1103/PhysRevLett.113.068102},
  url = {https://link.aps.org/doi/10.1103/PhysRevLett.113.068102}
}

@article{torre2026non,
  title={Non-Local Magic from the Entanglement Spectrum},
  author={Torre, Gianpaolo and Franchini, Fabio and Giampaolo, Salvatore Marco},
  journal={arXiv preprint arXiv:2607.07808},
  year={2026}
}

@article{huang2026intrinsic,
  title={Intrinsic spectral structure of bipartite nonlocal magic resource},
  author={Huang, Xiao and Chen, Guanhua and Yao, Yao},
  journal={arXiv preprint arXiv:2606.24368},
  year={2026}
}

@article{karjula2026ebbs,
  title={The ebbs and flows of quantum learning and sensing},
  author={Karjula, Matias and Ojanen, Teemu and Ala-Nissila, Tapio and Ivaki, Moein N},
  journal={arXiv preprint arXiv:2608.20155},
  year={2026}
}

@article{chang2019evolution,
  title={Evolution of entanglement spectra under generic quantum dynamics},
  author={Chang, Po-Yao and Chen, Xiao and Gopalakrishnan, Sarang and Pixley, JH},
  journal={Physical review letters},
  volume={123},
  number={19},
  pages={190602},
  year={2019},
  publisher={APS}
}

@article{van2024embezzlement,
  title={Embezzlement of entanglement, quantum fields, and the classification of von Neumann algebras},
  author={van Luijk, Lauritz and Stottmeister, Alexander and Werner, Reinhard F and Wilming, Henrik},
  journal={arXiv preprint arXiv:2401.07299},
  year={2024}
}

@article{PhysRevA.90.042331,
  title = {Characteristics of universal embezzling families},
  author = {Leung, Debbie and Wang, Bingjie},
  journal = {Phys. Rev. A},
  volume = {90},
  issue = {4},
  pages = {042331},
  numpages = {8},
  year = {2014},
  month = {Oct},
  publisher = {American Physical Society},
  doi = {10.1103/PhysRevA.90.042331},
  url = {https://link.aps.org/doi/10.1103/PhysRevA.90.042331}
}

@article{popescu2006entanglement,
  title={Entanglement and the foundations of statistical mechanics},
  author={Popescu, Sandu and Short, Anthony J and Winter, Andreas},
  journal={Nature Physics},
  volume={2},
  number={11},
  pages={754--758},
  year={2006},
  publisher={Nature Publishing Group UK London}
}

@article{hayden2006aspects,
  title={Aspects of generic entanglement},
  author={Hayden, Patrick and Leung, Debbie W and Winter, Andreas},
  journal={Communications in mathematical physics},
  volume={265},
  number={1},
  pages={95--117},
  year={2006},
  publisher={Springer}
}

@article{RevModPhys.96.025005,
  title = {Catalysis in quantum information theory},
  author = {Lipka-Bartosik, Patryk and Wilming, Henrik and Ng, Nelly H. Y.},
  journal = {Rev. Mod. Phys.},
  volume = {96},
  issue = {2},
  pages = {025005},
  numpages = {56},
  year = {2024},
  month = {Jun},
  publisher = {American Physical Society},
  doi = {10.1103/RevModPhys.96.025005},
  url = {https://link.aps.org/doi/10.1103/RevModPhys.96.025005}
}

@article{van2025multipartite,
  title={Multipartite embezzlement of entanglement},
  author={van Luijk, Lauritz and Stottmeister, Alexander and Wilming, Henrik},
  journal={Quantum},
  volume={9},
  pages={1818},
  year={2025},
  publisher={Verein zur F{\"o}rderung des Open Access Publizierens in den Quantenwissenschaften}
}

@article{fisher2023random,
  title={Random quantum circuits},
  author={Fisher, Matthew PA and Khemani, Vedika and Nahum, Adam and Vijay, Sagar},
  journal={Annual Review of Condensed Matter Physics},
  volume={14},
  number={1},
  pages={335--379},
  year={2023},
  publisher={Annual Reviews}
}

@article{PhysRevA.71.022316,
  title = {Universal quantum computation with ideal Clifford gates and noisy ancillas},
  author = {Bravyi, Sergey and Kitaev, Alexei},
  journal = {Phys. Rev. A},
  volume = {71},
  issue = {2},
  pages = {022316},
  numpages = {14},
  year = {2005},
  month = {Feb},
  publisher = {American Physical Society},
  doi = {10.1103/},
  url = {https://link.aps.org/doi/10.1103/PhysRevA.71.022316}
}

@article{kaufman2016quantum,
  title={Quantum thermalization through entanglement in an isolated many-body system},
  author={Kaufman, Adam M and Tai, M Eric and Lukin, Alexander and Rispoli, Matthew and Schittko, Robert and Preiss, Philipp M and Greiner, Markus},
  journal={Science},
  volume={353},
  number={6301},
  pages={794--800},
  year={2016},
  publisher={American Association for the Advancement of Science}
}

@article{collura2026nonlocal,
  title={Nonlocal nonstabilizerness in free fermion models},
  author={Collura, Mario and B{\'e}ri, Benjamin and Tirrito, Emanuele},
  journal={arXiv preprint arXiv:2604.27055},
  year={2026}
}

@article{iannotti2026non,
  title={Non-local magic resources for fermionic gaussian states},
  author={Iannotti, Daniele and Magni, Beatrice and Cioli, Riccardo and Hamma, Alioscia and Turkeshi, Xhek},
  journal={arXiv preprint arXiv:2604.27049},
  year={2026}
}

@article{nahum2017quantum,
  title={Quantum entanglement growth under random unitary dynamics},
  author={Nahum, Adam and Ruhman, Jonathan and Vijay, Sagar and Haah, Jeongwan},
  journal={Physical Review X},
  volume={7},
  number={3},
  pages={031016},
  year={2017},
  publisher={APS}
}

@article{zhang2026revealing,
  title={Revealing Entanglement-Growth Mechanisms through the Magic Barrier},
  author={Zhang, Lv and Zhang, Shi-Xin and Fan, Heng and Liu, Shuo},
  journal={arXiv preprint arXiv:2607.09875},
  year={2026}
}

@article{zanoni2024complete,
  title={Complete characterization of entanglement embezzlement},
  author={Zanoni, Elia and Theurer, Thomas and Gour, Gilad},
  journal={Quantum},
  volume={8},
  pages={1368},
  year={2024},
  publisher={Verein zur F{\"o}rderung des Open Access Publizierens in den Quantenwissenschaften}
}

@article{PhysRevLett.133.261602,
  title = {Relativistic Quantum Fields Are Universal Entanglement Embezzlers},
  author = {van Luijk, Lauritz and Stottmeister, Alexander and Werner, Reinhard F. and Wilming, Henrik},
  journal = {Phys. Rev. Lett.},
  volume = {133},
  issue = {26},
  pages = {261602},
  numpages = {8},
  year = {2024},
  month = {Dec},
  publisher = {American Physical Society},
  doi = {10.1103/PhysRevLett.133.261602},
  url = {https://link.aps.org/doi/10.1103/PhysRevLett.133.261602}
}

@article{viscardi2026non,
  title={Non-local Magic: closed-form solution and equivalence with magic of purification},
  author={Viscardi, Michele and Leone, Lorenzo and Hamma, Alioscia},
  journal={arXiv preprint arXiv:2609.04119},
  year={2026}
}

@article{liu2026nonlocal,
  title={Nonlocal nonstabilizerness for slightly entangled quantum many-body states},
  author={Liu, Lei-Yi-Nan and Cui, Jian},
  journal={arXiv preprint arXiv:2607.10714},
  year={2026}
}

@article{liu2026entanglement,
  title={Entanglement, anti-flatness, and nonlocal nonstabilizerness: a unified perspective from entanglement spectrum},
  author={Liu, Lei-Yi-Nan and Cui, Jian},
  journal={arXiv preprint arXiv:2609.01993},
  year={2026}
}

@article{franchini2026schmidt,
  title={Schmidt-Gauge Non-Local Magic: Representation, Optimality, and Mathematical Properties},
  author={Franchini, Fabio and Giampaolo, Salvatore Marco},
  journal={arXiv preprint arXiv:2608.02745},
  year={2026}
}

@article{sommers2004statistical,
  title={Statistical properties of random density matrices},
  author={Sommers, Hans-J{\"u}rgen and {\.Z}yczkowski, Karol},
  journal={Journal of Physics A: Mathematical and General},
  volume={37},
  number={35},
  pages={8457--8466},
  year={2004}
}

@article{vivo2016random,
  title={Random pure states: Quantifying bipartite entanglement beyond the linear statistics},
  author={Vivo, Pierpaolo and Pato, Mauricio P and Oshanin, Gleb},
  journal={Physical Review E},
  volume={93},
  number={5},
  pages={052106},
  year={2016},
  publisher={APS}
}

@article{pfyl-hwf2,
  title = {Complexity of entanglement embezzlement},
  author = {Schwartzman, Tal},
  journal = {Phys. Rev. A},
  volume = {112},
  issue = {1},
  pages = {012415},
  numpages = {10},
  year = {2025},
  month = {Jul},
  publisher = {American Physical Society},
  doi = {10.1103/pfyl-hwf2},
  url = {https://link.aps.org/doi/10.1103/pfyl-hwf2}
}

@article{PhysRevLett.83.3566,
  title = {Entanglement-Assisted Local Manipulation of Pure Quantum States},
  author = {Jonathan, Daniel and Plenio, Martin B.},
  journal = {Phys. Rev. Lett.},
  volume = {83},
  issue = {17},
  pages = {3566--3569},
  numpages = {0},
  year = {1999},
  month = {Oct},
  publisher = {American Physical Society},
  doi = {10.1103/PhysRevLett.83.3566},
  url = {https://link.aps.org/doi/10.1103/PhysRevLett.83.3566}
}

@article{PhysRevA.62.012304,
  title = {Approximate transformations and robust manipulation of bipartite pure-state entanglement},
  author = {Vidal, Guifr\'e and Jonathan, Daniel and Nielsen, M. A.},
  journal = {Phys. Rev. A},
  volume = {62},
  issue = {1},
  pages = {012304},
  numpages = {10},
  year = {2000},
  month = {Jun},
  publisher = {American Physical Society},
  doi = {10.1103/PhysRevA.62.012304},
  url = {https://link.aps.org/doi/10.1103/PhysRevA.62.012304}
}

@article{zhang2026entanglement,
  title={Entanglement Growth as Transport Across Schmidt Scales},
  author={Zhang, Shi-Xin and Liu, Shuo and Chen, Yu-Qin},
  journal={arXiv preprint arXiv:2609.06643},
  year={2026}
}

@article{PhysRevLett.127.150503,
  title = {Catalytic Transformations of Pure Entangled States},
  author = {Kondra, Tulja Varun and Datta, Chandan and Streltsov, Alexander},
  journal = {Phys. Rev. Lett.},
  volume = {127},
  issue = {15},
  pages = {150503},
  numpages = {6},
  year = {2021},
  month = {Oct},
  publisher = {American Physical Society},
  doi = {10.1103/PhysRevLett.127.150503},
  url = {https://link.aps.org/doi/10.1103/PhysRevLett.127.150503}
}
\end{document}